\documentclass[11pt, reprint,twocolumn,notitlepage,longbibliography,superscriptaddress]{revtex4-1}

\usepackage{graphicx}
\usepackage{amsmath} 
\usepackage{amsfonts}
\usepackage{bbm}
\usepackage{braket}
\usepackage{caption}
\usepackage{graphicx}
\usepackage{float}
\usepackage{subcaption}
\usepackage{url}
\usepackage{qcircuit}
\usepackage{amssymb}
\usepackage{qcircuit}
\usepackage{caption}

\begin{document}

\title{Controlling error orientation to improve quantum algorithm success rates}

\author{Daniel C. Murphy}
\affiliation{School of Physics, Georgia Institute of Technology, Atlanta, Georgia 30332, USA}

\author{Kenneth R. Brown}
\affiliation{Departments of Electrical and Computer Engineering, Chemistry, and Physics, Duke University, Durham, NC 27708, USA}
\affiliation{Schools of Chemistry and Computational Science Engineering, Georgia Institute of Technology, Atlanta, Georgia 30332, USA}
\affiliation{School of Physics, Georgia Institute of Technology, Atlanta, Georgia 30332, USA}
\email{ken.brown@duke.edu}

\begin{abstract}
The success probability of a quantum algorithm constructed from noisy quantum gates cannot be accurately predicted from single parameter metrics that compare noisy and ideal gates. We illustrate this concept by examining a system with coherent errors and comparing algorithm success rates for different choices of two-qubit gates that are constructed from composite pulse sequences, where the residual gate errors are related by a unitary transformation. As a result, all of the sequences have the same error relative to the ideal gate under any distance measure that is invariant under unitary transformations. However, the circuit success can vary dramatically by choosing error orientations that do not affect the final outcome and error orientations that cancel between conjugate controlled-nots, as demonstrated here with Clifford circuits, compiled Toffoli gates, and quantum simulation algorithms. The results point to the utility of  both minimizing the error and optimzing the error direction and also to the advantages of using multiple control sequences for the same gate type within a single algorithm.            
\end{abstract}

\maketitle

\section{Introduction}

The success probability of quantum algorithms depends on the quality of underlying operations. In the near-term, noisy quantum gates will be a feature of all quantum hardware \cite{PreskillQuantum2018}. For every quantum gate there are multiple control sequences that apply the same gate but yield different errors, even in the same physical system. This control flexibility provides multiple methods for improving algorithms in the face of errors  For stochastic errors, if one can control the weight and type of stochastic errors, it is possible to run a series of imperfect quantum circuits and extrapolate to the expected value of a perfect quantum circuit \cite{LiPRX2017, TemmePRL2017, DumitrescuPRL2018, KandalaArXiv2018}.  For errors described by unitary transformations,  quantum control methods are available to generate gates with smaller and smaller error including dynamic decoupling \cite{Hahn1950,ViolaPRL1999,BiercukJPB2011,QuirozPRA2013,QIQIP2017}, composite pulses \cite{LevittPNMRS1986, WimperisJMR1994, BrownPRA2004, TorosovPRA2011, BandoJPSJ2013, MerrillACP2014, LowPRA2014, LowPRX2016, CohenPRA2016}, and optimal control \cite{MontangeroPRL2007, GraceJMO2007,SchulteJPB2011, MachnesPRL2018}.

Typical discussions of errors focus on scalar metrics such as the fidelity or the diamond norm \cite{SandersNJP2016, KuengPRL2016, IyerQST2018} but error operators also have an orientation in the space of operators. A quantum algorithm starts with state preparation and ends with measurement and both the state preparation and the measurement break the symmetry of errors. An ideal case would be where the error of every gate maps to an error at the end of the circuit which commutes with the measurement and is therefore harmless.  To achieve this goal, we would need a method to control the error direction of each gate and to understand how it propagates through the circuit.

We show this goal is achievable in the context of two-qubit gates with systematic coherent errors.  First, we define our notation for unitary operations and our two fidelity measures: gate fidelity and circuit fidelity. Then we explain how a two-qubit gate composite pulse sequence allows us to transform what is initially a two-qubit error into a single-qubit error on either qubit and oriented in any direction. Next, using an implementation of the Bernstein-Vazirani algorithm \cite{BernsteinSTOC93} as an example, we show how we can trace the single qubit errors through a Clifford circuit to determine the optimal orientation for each gate. Finally, we show how this method can be used in non-Clifford circuits by taking advantage of common patterns of controlled-not gates in quantum algorithms.  This work shows that different implementations of the same two-qubit gate, which have the same fidelity and diamond norm, can lead to algorithmic failure probabilities that differ by orders of magnitude. 

\section{Notation, Error Model, and Fidelity Measures}

Following standard quantum information notation, we write the Pauli operators as $X$,$Y$, and $Z$. Other standard gates are the Hadamard, $H=\sqrt{1/2} (X+Z)$, the $\pi/8$ or $T$ gate, $T=\exp(i\pi/8)(\cos(\pi/8)I-i\sin(\pi/8)Z)$, and the controlled-not is a two qubit gate acting on the control qubit $c$ and the target qubit $t$, CNOT$(c,t)$=$\ket{0}\bra{0}_cI_t+\ket{1}\bra{1}_cX_t$ \cite{N&C}. 

We are primarily interested in unitary operations $U$ generated by a Hermitian operator $A$. Our typical notation will be \begin{eqnarray}
U_A(\theta)=\exp(-i\frac{\theta}{2}A)
\end{eqnarray}
and our convention is guided by the rotations of single qubits $U_Z(\theta)=R_z(\theta)$. It will be convenient to have parameterized operators $\Phi_{A_i,A_j}=\cos(\phi)A_i+\sin(\phi)A_j$.  

The error model considered in this paper are unknown but systematic and proportional errors in $\theta$. This is a common problem experimentally due to slow fluctuations in the value of the control field. It is convenient to introduce the following notation:
\begin{equation}
V_A(\theta)=U_A(\theta(1+\epsilon)),
\end{equation}
where $V_A(\theta)$ represents an attempt to apply $U_A(\theta)$ that is ruined by noise.

We quantify the fidelity of our $n$ qubit gates using the entanglement fidelity~\cite{SchumacherPRA1996} to compare our ideal gate $U_G$ and the applied gate $\tilde{U}_G$. Since we are only considering unitary errors, we can write  $ U_G^\dagger \tilde{U}_G$ as the gate associated error channel and the entanglement fidelity is
\begin{equation}
    \mathcal{F}_\mathrm{gate} = \left|\frac{tr[U_G^\dagger \tilde{U}_G]}{2^n}\right|^2
\end{equation}

To quantify our circuit fidelity, we compare the actual output with with the desired output. In many algorithms, the output register $O$ is only a fraction of all of the qubits involved in the circuit $C$. In this paper, we consider only algorithms where the ideal output is a pure state
\begin{equation}
\mathcal{F}_\mathrm{circuit} = Tr[\rho_C (\ket{\psi}\bra{\psi}_{O}\otimes I_{C-O})]
\end{equation}
where $\rho_C$ is the final state of all the qubits after the actual circuit and $\ket{\psi}_O$ is the ideal state on the output. 

The circuit fidelity can be much higher than the gate fidelity.  Consider a circuit which consists of prepare 0, apply $H$, and measure in the X basis.  The ideal output is $\ket{+}$. Consider two erroneous $H$ gates: $H_a=U_X(\epsilon)H$ and $H_b=U_Z(\epsilon)H$.  The gates have the same gate fidelity but $H_b$ has no impact on circuit fidelity,  
\begin{eqnarray}
\mathcal{F}_\mathrm{gate}(H_a) =
\mathcal{F}_\mathrm{gate}(H_b)=\mathcal{F}_\mathrm{circuit}(H_b) &=&\cos(\epsilon/2)^2 \\
\mathcal{F}_\mathrm{circuit}(H_a)&=&1
\end{eqnarray}

This trivial example shows that the orientation of the error can have a large effect on circuit fidelity and that any one parameter metric of gate error may be a poor predictor of algorithmic success. $H_a$ and $H_b$ have the same distance from $H$ and the same fidelity loss compared to $H$ for any metric that is invariant under unitary transformations.  In the remainder of the paper, we will show that this difference in $\mathcal{F}_\mathrm{circuit}$ for constant $\mathcal{F}_\mathrm{gates}$ can occur naturally by combining composite pulses and two-qubit gates with systematic errors.

\section{Composite pulses for Ising-type interactions}

In our model the only error is a systematic error in the implementation of two-qubit gates.  Each two-qubit gate is generated by an Ising-type Hamiltonian with the ideal two-qubit gate being $U_{XX}(\theta)=\exp(-i\theta/2 XX)$, where $X$, $Y$, and $Z$ represent the single qubit Pauli operators. Our error is that the non-ideal gate $V$ has a systematic error such that $V_{XX}(\theta)=U_{XX}(\theta(1+\epsilon))$. This error naturally occurs due to imprecise timing and slow fluctuations of the control fields. 

Jones showed that under the assumption of perfect single qubit gates the errors in $V$ can be mitigated using composite pulse sequences developed for single-qubit gates \cite{JonesPRA2003}.  Composite pulses on single-qubit gates relies on the algebra of su(2): $[A_i,A_j]=i\epsilon_{ijk}2A_k$.  For single qubits, $A_1=X$,$A_2=Y$, $A_3=Z$.  For our problem, we are given $A_1=XX$ and we are free to choose $A_2$ from any weight-2 Pauli operator that anticommutes with $XX$. Regardless of our choice, $A_3$, is a weight-1 Pauli operator that anticommutes with $XX$ \cite{TomitaNJP2010}.  In analogy with single-qubit operators: we denote $A_j$ as the operator and $a_j$ as the direction in the Lie algebra.

Since $A_1$ and $A_2$ are unitary and Hermitian, $A_1A_1=A_2A_2=I$, and anticommute $A_1A_2=-A_2A_1$, we have simple form for $U_{\Phi_{i,j}}(\theta)$,
\begin{eqnarray}
U_{\Phi_{i,j}}(\theta)=& &\cos(\theta/2)I\nonumber\\ &-& i\sin(\theta/2)\left(\cos(\phi_{i,j})A_i+\sin(\phi_{i,j})A_j\right)
\end{eqnarray}

For concreteness, we choose the composite pulse sequence SK1 \cite{BrownPRA2004}, 
\begin{eqnarray}
    SK1_{A_1}(\theta)&=&V_{-\Phi^{SK1}_{1,2}}(2\pi)V_{\Phi^{SK1}_{1,2}}(2\pi)V_{A_1}(\theta)\\
    &=&R_{A_3}(\beta\epsilon^2\theta^2)R_{A_1}(\theta)+O(\epsilon^3),
\end{eqnarray}
where $\phi^{SK1} = \cos^{-1}\left(-\frac{\theta}{4\pi}\right)$ and $\beta=4\pi^2\sin(\phi^{SK1})\cos(\phi^{SK1})$. The important feature of SK1 for this study is that the leading order of the error is a rotation about $A_3$ and this will hold for all fully compensating pulses that correct to $\epsilon^n$ where $n$ is even. 

\begin{figure}[t]
     \centering
    \includegraphics[width=6cm,scale=0.5]{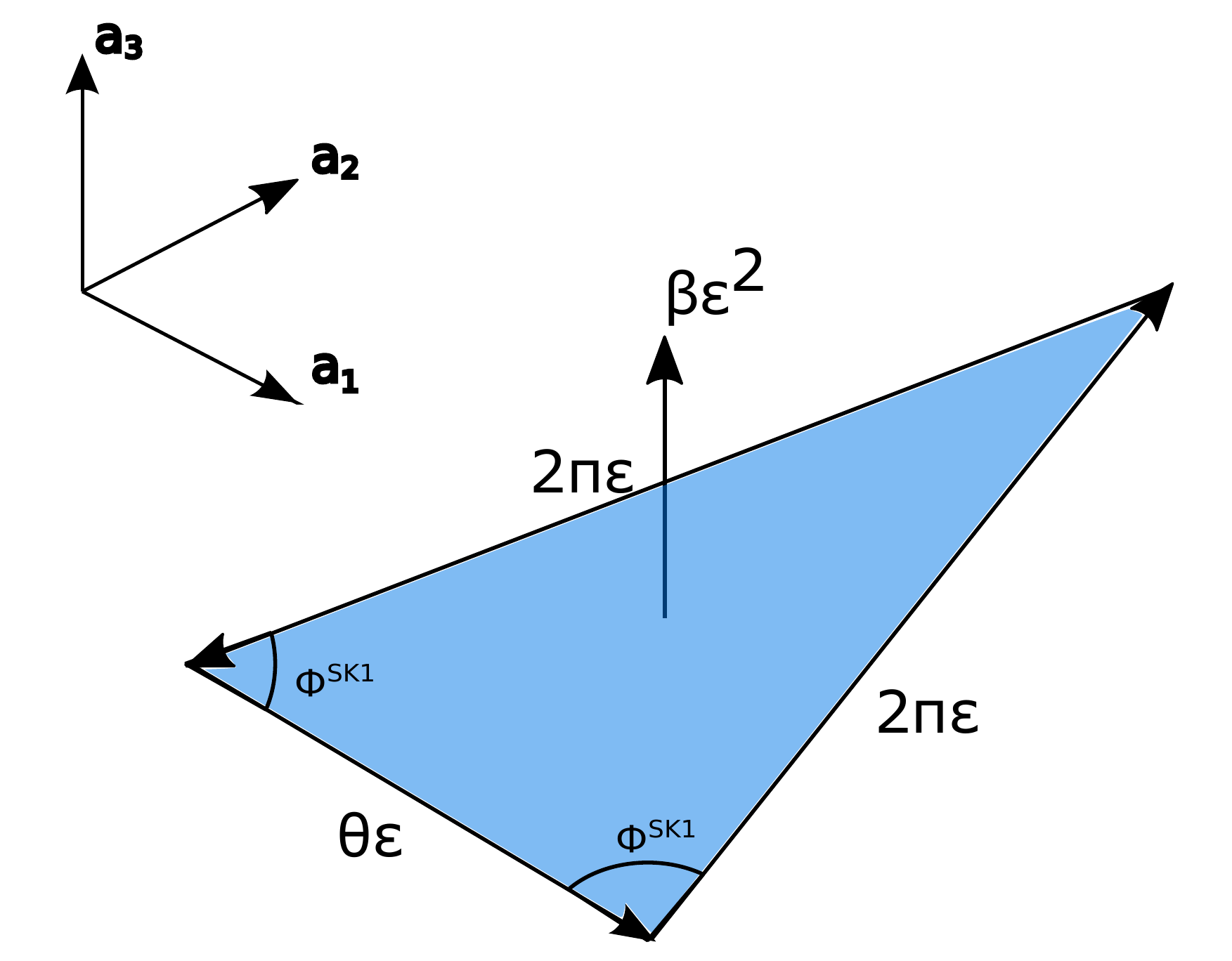}
     \caption{A visual of the SK1 pulse sequence. The dominant residual error points orthogonal to the plane of correction (in this figure $a_1$ - $a_2$). This second order rotation has an angle of rotation proportional to the pulse area, $\beta = 4\pi^2\sin(\phi^{SK1})\cos(\phi^{SK1})$}
     \label{fig:SK1Visual}
 \end{figure}

Given $A_1=XX$, the following choices of $A_3$ are possible $YI$,~$ZI$,~$IY$, and ~$IZ$. This transformation of a two-qubit error into a predominantly single-qubit error gives us a number of opportunities for error cancellation. We can also convert the error to an $XI$ or $IX$ error by conjugating the composite pulse that yield $Z$ errors with $\pi/2$ rotations about $YY$ at the cost of additional gates \cite{TomitaNJP2010}.

In the rest of the paper, we focus on CNOT gates.  These gates are constructed from rotations about $\pi/2$ around $XX$ and local gates \cite{DebnathNature2016, MaslovNJP2017}.  These local gates transform the residual errors to a choice of $ZI$, $XI$, $IZ$, and $IY$.  CNOT is its own inverse and conjugating the gate sequence yields an equivalent CNOT where the rotation error can be considered coming before the gate and will have the opposite sign. 

For completeness, we write out the pulse sequences that yield rotation errors after the gate around $XI$ and $YI$ on the control qubit and $IY$ on the target qubit:

\begin{widetext}
\begin{eqnarray}
W_2&=&U_{ZI}(-\pi/2)U_{YI}(-\pi/2)U_{IX}(-\pi/2)\\
W_1&=&U_{YI}(\pi/2)\\
\mathrm{CNOT}&=& W_2 V_{XX}(\pi/2) W_1  \\
\mathrm{CNOT}_{YI}&=&W_2U_{YI}(\phi^{SK1})V_{XX}(2\pi)U_{YI}(-2\phi^{SK1})V_{XX}(2\pi)U_{YI}(\phi^{SK1})V_{XX}(\pi/2)W_1\\
\mathrm{CNOT}_{IY}&=&W_2U_{IY}(\phi^{SK1})V_{XX}(2\pi)U_{IY}(-2\phi^{SK1})V_{XX}(2\pi)U_{IY}(\phi^{SK1})V_{XX}(\pi/2)W_1\\
\mathrm{CNOT}_{XI}&=&W_2U_{ZI}(-\phi^{SK1})V_{XX}(2\pi)U_{ZI}(2\phi^{SK1})V_{XX}(2\pi)U_{ZI}(-\phi^{SK1})V_{XX}(\pi/2)W_1\\
\mathrm{CNOT}_{-XI}&=&CNOT_{XI}^{\dagger}
\end{eqnarray}
\end{widetext}
in these sequences $\phi^{SK1}=\cos^-1(-1/8)$ and $V_A^\dagger(\theta)=V_A(-\theta)$.

\section{Strategies for reducing circuit error by controlling gate orientation}

We employ two strategies for reducing circuit error.  The first strategy notes that preparation and measurement sets a basis. Any erroneous unitary operation that preserves the basis will not change the measurement outcome and generates only a global phase on the prepared state.  The second strategy notes that CNOTs often form conjugate pairs around other unitary operations. We can then carefully choose the gate sequence so the leading order residual coherent error is on the control qubit and is cancelled between the paired CNOTs.

\subsection{Tracing errors through the circuit}

After applying the composite pulse, the residual error is a single qubit rotation.  We can track how the Pauli operator transforms from the error location to measurement to determine which Pauli error we should choose.  This generally can be quite hard but if we restrict ourselves to Clifford circuits we can take advantage of fault-path tracing methods developed for quantum error correction \cite{AliferisQIC2006,JanardanQIP2016,MillerArXiv2018}

As an example consider an implementation of the Bernstein-Vazirani circuit commonly used in experiments \cite{FallekNJP2016, LinkePNAS2017}.  The final measurement is in the Z basis, between the CNOT and the Z measurement is a Hadamard gate that changes X to Z.  We then simply choose the composite pulse that yield an error which is primarily an X rotation on the control qubit. 


    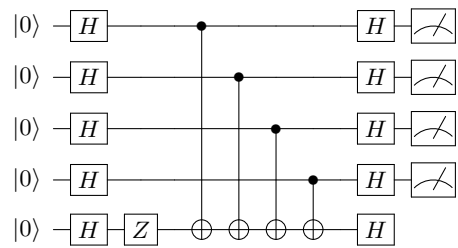
\begin{figure}[t]
        \centering
        \[ \Qcircuit @C=.7em @R=.7em { 
             \lstick{\ket{0}} & \gate{H} & \qw & \qw & \ctrl{4} & \qw & \qw & \qw & \qw & \gate{H} & \meter      \\
             \lstick{\ket{0}} & \gate{H} & \qw & \qw & \qw & \ctrl{3} & \qw & \qw & \qw & \gate{H} & \meter      \\
             \lstick{\ket{0}} & \gate{H} & \qw & \qw & \qw & \qw & \ctrl{2} & \qw & \qw & \gate{H} & \meter      \\
             \lstick{\ket{0}} & \gate{H} & \qw & \qw & \qw & \qw & \qw & \ctrl{1} & \qw & \gate{H} & \meter      \\
             \lstick{\ket{0}} & \gate{H} & \gate{Z}  & \qw & \targ & \targ & \targ & \targ & \qw & \gate{H}
         } \]

\caption{A circuit for implementing the Bernstein-Vazarani for $f(\mathbf{x}) = \mathbf{a}\cdot\mathbf{x}$ given $\mathbf{a} = 1111$.}
        \label{fig:BVCircuit}
    \end{figure}
    \begin{figure}[h]
        \centering
        \includegraphics[width=0.5\textwidth]{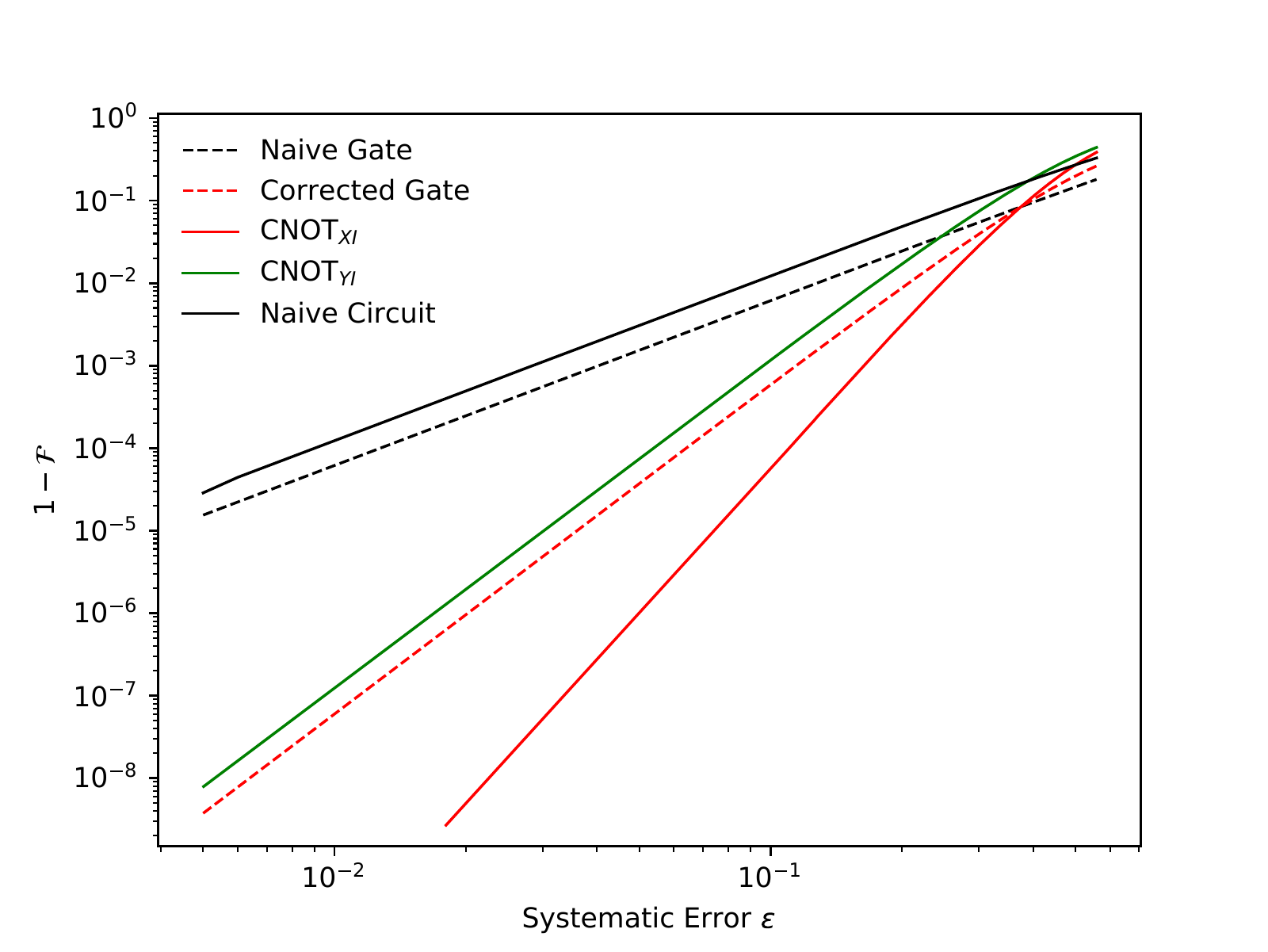}
        \caption{Comparison of $1-\mathcal{F}_\mathrm{circuit}$  for the 4-bit Bernstein-Vazarani circuit for different CNOT implementations as a function of the systematic error $\epsilon$ (solid lines). The underlying $1-\mathcal{F}_\mathrm{gate}$ are plotted for the different implementations of CNOT (dashed lines). Although the corrected CNOTs ( CNOT$_{XI}$ and CNOT$_{YI}$) have the same $\mathcal{F}_\mathrm{gate}$, they result in drastically different $\mathcal{F}_\mathrm{circuit}$.}
        \label{fig:BVResults}
    \end{figure}
   

Regardless of the error orientation the SK1 gates have the same fidelity and diamond distance.  They both yield an improvement in the circuit fidelity relative to a naive gate but the SK1 gate where the error is oriented along $X$ significantly outperforms the alternative choice (Fig. \ref{fig:BVResults}). For the naive gate and the poorly chosen composite pulse gate, the circuit error and gate error decreases by the same power of $\epsilon$.  By choosing the composite pulse sequence that leads to the error vanishing on the measurement the circuit error decreases as $\epsilon^6$ compared to gate errors that are decreasing as $\epsilon^4$. Tracing error can even work in circuits not made from Clifford gates but it is only practical for circuits of limited depth and may not yield as dramatic gains.  In these cases, it may be more valuable to construct more robust gate sequences locally within the circuit. 

\subsection{Canceling errors with CNOT conjugation}
 A common pattern in quantum algorithms is an operation placed between the target of two CNOTs. The control bit is not modified during this operation and the by aligning the errors to cancel on the control bit, we can again reduce the circuit error. We present two examples of this procedure in practice: the Toffoli gate and a quantum simualtion algorithm.
 
 \subsubsection{Toffoli Gate}
 
\begin{figure}[t]
     \centering
    \includegraphics[width=0.5\textwidth]{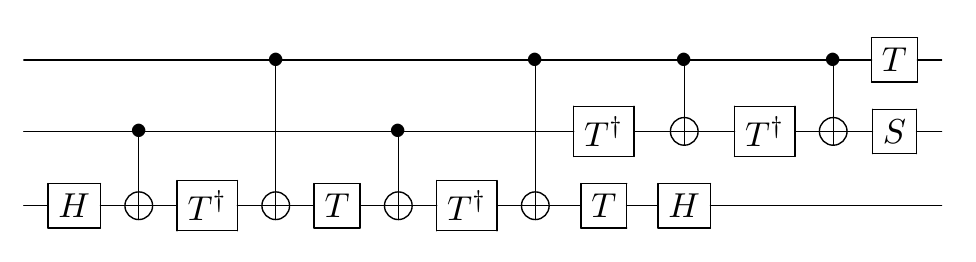}
     
\caption{A circuit for implementing the Toffoli gate from single qubit gates and CNOT gates. Note that the 6 CNOT gates come in 3 conjugate pairs.}
\label{fig:ToffoliCircuit}
\end{figure}

 Consider the canonical circuit for implementing a Toffoli gate from CNOTs (Fig. \ref{fig:ToffoliCircuit}), T gates, and Hadamards \cite{N&C}. The $6$  CNOTs form $3$ pairs of conjugated CNOTs. 
 
\begin{figure}[h]
\centering
    \includegraphics[width=0.5\textwidth]{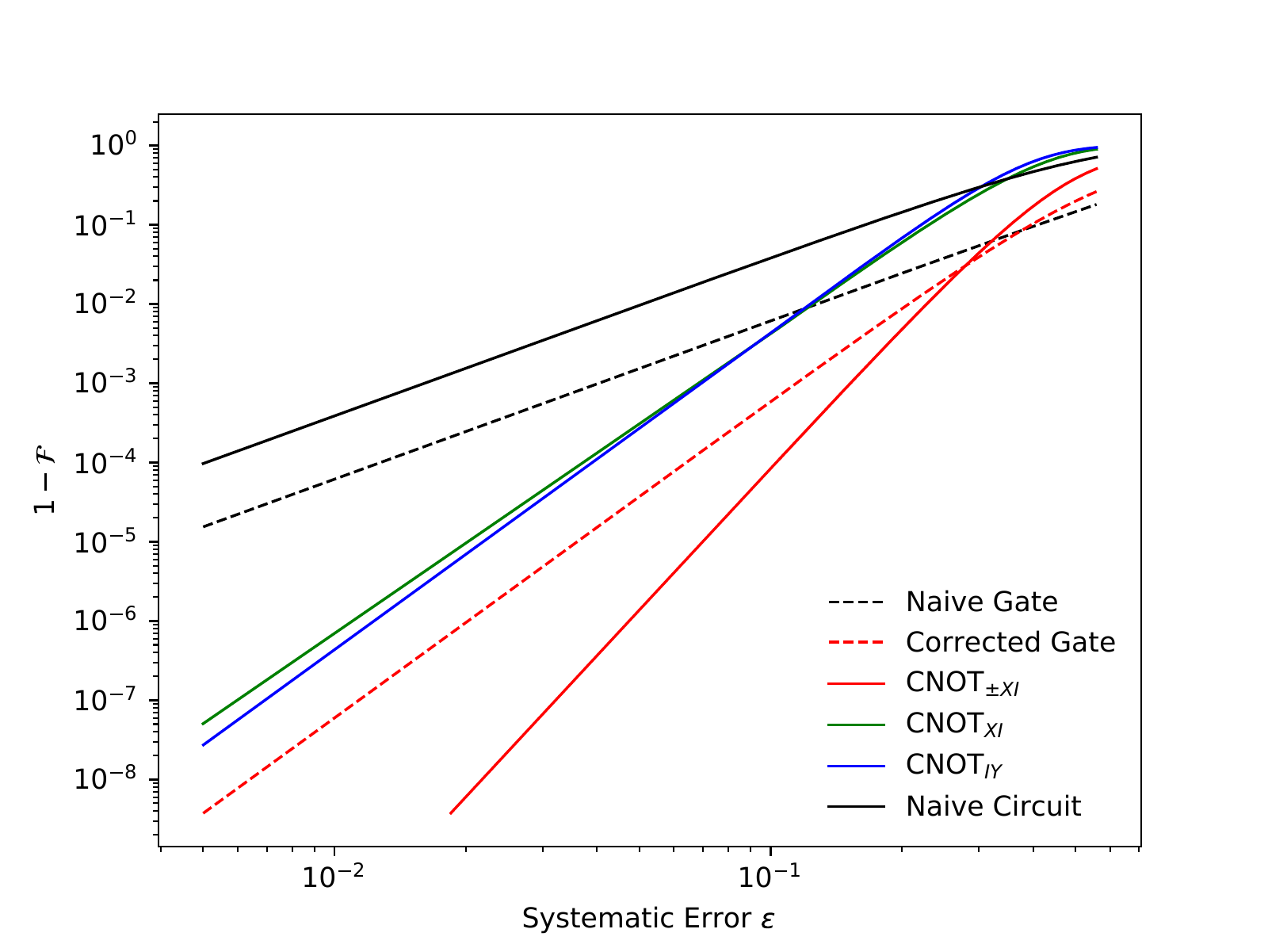}
\caption{Comparison of of 1-$\mathcal{F}_\mathrm{gate}$ of the Toffoli gate as a function of the systematic error $\epsilon$ for different CNOT pulse sequences (solid lines). The underlying gate errors of the CNOT sequences making up the Toffoli (Fig. \ref{fig:ToffoliCircuit}) are also presented (dashed lines).  Corrected pulse sequences improve gate fidelity for small $\epsilon$ over the naive circuit, but a strategy of cancelling errors on the target qubit by conjugate pairs yields the lowest loss in Toffoli gate fidelity and a fundamentally improved scaling.}\label{fig:ToffoliInfidelity}
 \end{figure}
 
 We compare four choices of gates: naive gates, CNOT, SK1 with errors on the control bit, CNOT$_{XI}$, SK1 with errors on the target bit, CNOT$_{IY}$, and SK1 with conjugate error cancellation on the control bit, CNOT$_{\pm XI}$ (Fig. \ref{fig:ToffoliInfidelity}). Using naive CNOTs, the error of the Toffoli gate is always worse than the error of the CNOTs. This is not surprising, since the Toffoli is composed of 6 CNOTs. This remains true for CNOT$_{IY}$ and CNOT$_{XI}$, which improve the overall fidelity relative to CNOT for small errors. Using CNOT$_{\pm XI}$ to cancel errors on conjugate CNOT pairs, we find the unexpected behavior that the Toffoli gate fidelity loss is less than the underlying CNOT error, showing the value of this strategy for coherent error cancellation.

\section{Quantum Simulation and Phase Estimation} 

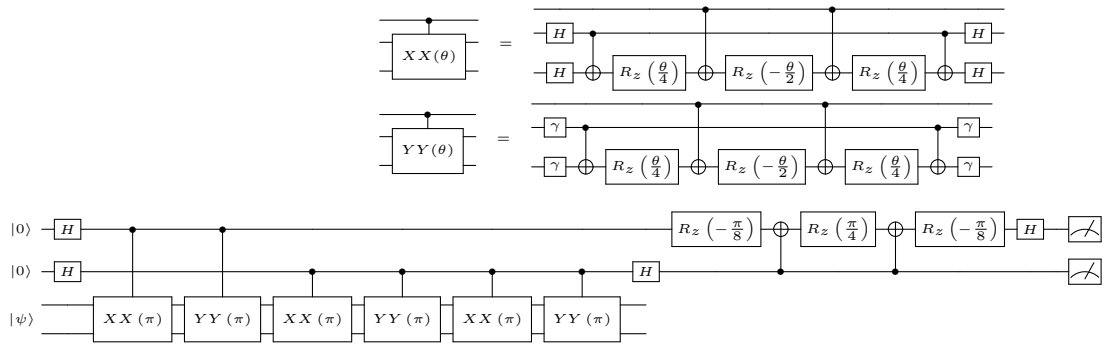
\begin{figure*}[t]
    \tiny
    \begin{subfigure}[t]{0.3\textwidth}
        \centering
        \begin{tabular}{ccc}
            \begin{tabular}{c}
               \Qcircuit @C= .7 em @R=.7 em { 
                    & \ctrl{1}                  & \qw \\
                    & \multigate{1}{XX(\theta)} & \qw \\
                    & \ghost{XX(\theta)}        & \qw 
                }
            \end{tabular}& 
            \begin{tabular}{c} = \end{tabular} & 
            \begin{tabular}{c}
                \Qcircuit @C= .7 em @R=.7 em { 
                    & \qw & \qw & \qw & \ctrl{2} & \qw & \ctrl{2} & \qw & \qw & \qw & \qw \\
                    & \gate{H} & \ctrl{1} &  \qw & \qw & \qw & \qw &\qw & \ctrl{1} & \gate{H} & \qw\\
                    & \gate{H} & \targ & \gate{R_z\left(\frac{\theta}{4}\right)} & \targ & \gate{R_z\left(-\frac{\theta}{2}\right)} & \targ & \gate{R_z\left(\frac{\theta}{4}\right)} & \targ & \gate{H} & \qw
                }
            \end{tabular}
        \end{tabular}
        
        \begin{tabular}{ccc}
            \begin{tabular}{c}
                \Qcircuit @C= .7 em @R=.7 em { 
                    & \ctrl{1}                  & \qw \\
                    & \multigate{1}{YY(\theta)} & \qw \\
                    & \ghost{YY(\theta)}        & \qw 
                }
            \end{tabular}& 
            \begin{tabular}{c} = \end{tabular} & 
            \begin{tabular}{c}
                \Qcircuit @C= .7 em @R=.7 em { 
                    & \qw & \qw & \qw & \ctrl{2} & \qw & \ctrl{2} & \qw & \qw & \qw & \qw \\
                    & \gate{\gamma} & \ctrl{1} &  \qw & \qw & \qw & \qw &\qw & \ctrl{1} & \gate{\gamma} & \qw\\
                    & \gate{\gamma} & \targ & \gate{R_z\left(\frac{\theta}{4}\right)} & \targ & \gate{R_z\left(-\frac{\theta}{2}\right)} & \targ & \gate{R_z\left(\frac{\theta}{4}\right)} & \targ & \gate{\gamma} & \qw
                }
            \end{tabular}
        \end{tabular}
         \vspace{0.3 cm}
        
        \label{subfig:Rots}
    \end{subfigure}
   
    \begin{subfigure}[t]{0.3\textwidth}
        \Qcircuit @C= .7 em @R=.7 em { 
                    \lstick{\ket{0}} & \gate{H} & \ctrl{2} & \ctrl{2} & \qw & \qw & \qw & \qw & \qw & \gate{R_z\left(-\frac{\pi}{8}\right)} & \targ & \gate{R_z\left(\frac{\pi}{4}\right)} & \targ & \gate{R_z\left(-\frac{\pi}{8}\right)} & \gate{H} &  \qw & \meter\\
                    \lstick{\ket{0}} & \gate{H} & \qw & \qw & \ctrl{1} & \ctrl{1} & \ctrl{1} & \ctrl{1} & \gate{H} & \qw &  \ctrl{-1} & \qw & \ctrl{-1} & \qw & \qw &\qw  &\meter \\
                    \lstick{} & \qw & \multigate{1}{XX\left(\pi\right)} & \multigate{1}{YY\left(\pi\right)} & \multigate{1}{XX\left(\pi\right)} & \multigate{1}{YY\left(\pi\right)} & \multigate{1}{XX\left(\pi\right)} & \multigate{1}{YY\left(\pi\right)} & \qw\\
                    \lstick{} & \qw & \ghost{XX\left(\pi\right)} & \ghost{YY\left(\pi\right)} & \ghost{XX\left(\pi\right)} & \ghost{YY\left(\pi\right)} & \ghost{XX\left(\pi\right)} & \ghost{YY\left(\pi\right)} & \qw
                     \inputgrouph{3}{4}{.8em}{\ket{\psi}}{1.1em}\\
                }
        
        \label{subfig:PEA}
    \end{subfigure}

    \caption{The phase estimation circuit which simulates the Hamiltonian $XX + YY$. There is a symmetry for all of the CNOT gates within this circuit. All of the CNOT gates in this circuit have a neighboring CNOT gate, which acts on the same control such that no unitary acts on the control qubit in between them.  By ensuring the implementation of these neighboring CNOT pairs are the conjugate of each other with the dominant error placed on the control qubit,  the dominant residual error from the CNOT pair will become identity. This technique can be applied to any controlled rotation. The state $\ket{\psi}$ on which phase estimation is performed was chosen as the ground-state of $H$, $\frac{\ket{10}-\ket{01}}{\sqrt{2}}$}
    \label{fig:PEACircuit}
\end{figure*}

\begin{figure}[h]
    \centering
    \includegraphics[width=10cm,scale=0.5]{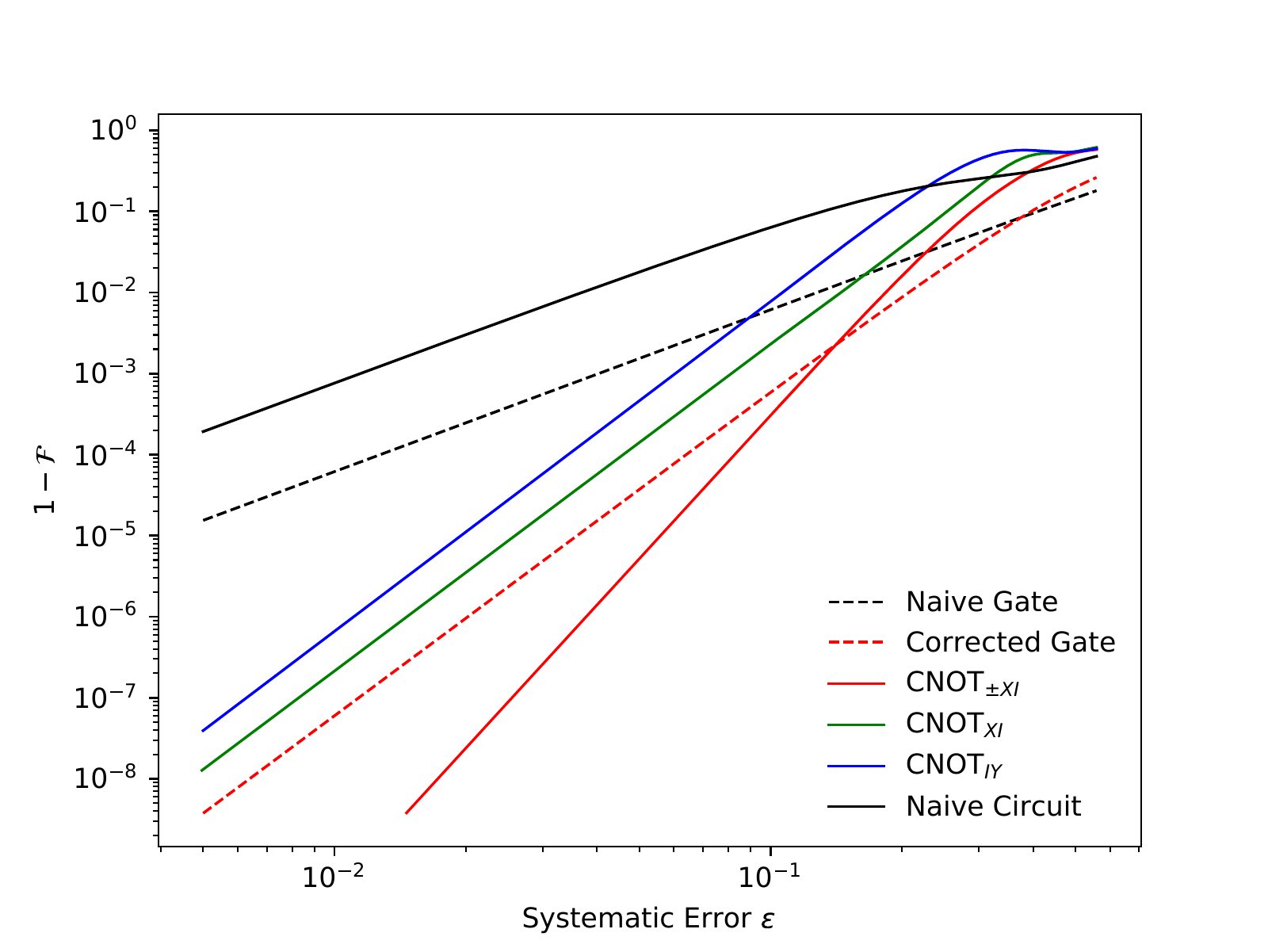}
    \caption{Comparison of $1-\mathcal{F}_\mathrm{circuit}$ for the quantum simulation circuit \ref{fig:PEACircuit} for different CNOT gates versus the systematic error parameter $\epsilon$ (solid lines). The underlying CNOT errors, $1-\mathcal{F}_\text{gate}$, for SK1 corrected and uncorrected CNOT gates are also plotted (dashed) lines. The best error suppression occurs when the corrected gates are chosen as conjugate pairs (CNOT$_{\pm XI}$). }
    \label{fig:PEAInfidelity}
 
\end{figure}
 
 Quantum simulation on a quantum computer requires generating Hamiltonian dynamics with multi-qubit operators.  Whether this is performed by Trotter decomposition \cite{LloydScience1996} or Taylor series expansions \cite{BerryPRL2015} or quantum signal processing \cite{LowPRL2017}, the underlying methods involve an array of conjugate CNOTs that transform single qubit operators into multiple qubit operators.  These conjugations of operators allow us to cancel the coherent errors using CNOT pairs as we did for the Toffoli gate.
 
 Quantum simulation is a subroutine in a phase estimation algorithm for finding the eignevalues of a Hamiltonian \cite{KitaevPEA, AbramsPRL1999}. The phase estimation requires the inverse QFT which is constructed from increasingly fine controlled rotations.  These controlled rotations can also be constructed from single qubit rotations and conjugate CNOTs.  Again this yields an opportunity to cancel coherent errors.

We demonstrate our method using a 4 qubit implementation of phase estimation to determine the eigenvalue of a two-qubit spin Hamiltonian. The $XY$ spin Hamiltonian $H_{XY}=\sum_{k,j} h_{k,j}(X_kX_j+Y_kY_j)$ has a number of applications in nuclear physics and condensed matter physics. Quantum simulations of this Hamiltonian with single qubit energy terms have been performed using nuclear magnetic resonance \cite{BrownPRL2006, PengPRL2010}, trapped ions \cite{LanyonScience2011,RichermeNature2014}, and superconducting ~\cite{SalathePRX2015,DumitrescuPRL2018}.   We have chosen the parameters such that the phase is exact for two qubit phase estimation. This example already shows the key opportunity for cancellation in the quantum simulation and in the quantum Fourier transform (Fig. \ref{fig:PEAInfidelity}). We allow for perfect state simulation in this case and note that state preparation is often part of the circuit where CNOTs do not appear in conjugate form.  In many cases the first strategy of attempting to cancel the errors on initialization instead of final measurement will be useful.
 
First, comparing SK1 sequences CNOT$_{XI}$ and CNOT$_{IY}$, we see in Fig.  \ref{fig:PEAInfidelity} that the error on the control line yields a lower circuit error for small errors and for both circuits the error is reduced as $\epsilon^2$ following the gate fidelity reduction of $\epsilon^2$.  The circuit with conjugate CNOTs (CNOT$_{\pm XI}$)  yields a circuit error that scales as $\epsilon^6$ and shows again that for small coherent error that the circuit fidelity is greater than the gate fidelity.

\section{Conclusion}

We have shown that controlling the orientation of errors greatly improves algorithm performance using a simple model of overrotation in two-qubit gates.  This work again emphasizes that single parameter characterizations of gate quality are limited predictors of circuit success \cite{IyerQST2018}. The SK1 gates considered here all have exactly the same fidelity and diamond norm relative to the ideal gate but remarkably different performance in the  context of the circuits. As a result, for any quantum control procedure one should optimize not only for error but also the orientation of the error. 

For higher-order composite pulse sequences where the residual error is orthogonal to the gate, our approach can be applied directly.  This is the case for amplitude errors when the pulse sequence has an even-order residual error \cite{LowPRA2014}.  For odd order residual errors, for example, BB1\cite{WimperisJMR1994}, the error is not orthogonal to the gate and the controlling orientation will have only a limited effect.  For numerical optimal control, error orientation can be added to the cost function for optimization. We generally expect that tuning conjugate pairs of CNOTs can reduce the effect of errors even when the error has only partial unitarity  \cite{WallmanNJP2015} based on modeling the cancellation of coherent errors in quantum error correcting protocols \cite{DebroyArXiv2018}.

Composite pulse sequences have been successful for single-qubit gates \cite{ShappertNJP2013,MerrillPRA2014,MountPRA2015,FallekNJP2016,BlumeKohoutNatComm2017}, but are generally unused for two qubit gates. The reason is that two-qubit gates typically have longer gate times than single qubit gates, which results in more stochastic noise per gate. The $SK1$ sequence studied here for $U_A(\pi/2)$ is 9 times longer than the naive gate and as a consequence stochastic errors will accumulate. Faster two-qubit gates or weaker stochastic errors are necessary for the sequences discussed here to be useful in the laboratory.

Experimentalists have access to a wide variety of pulse sequences that generates nominally the same gate. The broad principles of cancelling the errors locally or erasing them in measurement or state preparation can be used to help determine which gate sequence to choose.  These simple rules can help increase the length of quantum computations in this era of noisy quantum gates. 

\acknowledgements

The authors thank Michael Newman, Pak Hong Leung, and Leonardo Andreta de Castro for helpful conversations.  This work was supported
by the ARO MURI on Modular
Quantum Systems W911NF-16-1-0349, NSF Expeditions in Computing award 1730104, and the DOE ASCR award DE-SC0019294.

\bibliographystyle{apsrev}

\end{document}